\documentclass[aps,prx,reprint,showpacs,superscriptaddress,floatfix]{revtex4-2}
\usepackage[utf8]{inputenc}
\usepackage{graphicx}
\usepackage{booktabs}
\usepackage{tabularx}
\usepackage{hyperref}
\hypersetup{
colorlinks=true,final=true,
        linkcolor=blue,
        citecolor=blue,
        filecolor=blue,
        urlcolor=blue,
}

\newcommand{\lno}{{L}a$_{5}${N}i$_{3}${O}$_{11}$}
\newcommand{\ef}{$\varepsilon_{\mathrm{F}}$}
\newcommand{\edit}[1]{\textcolor{black}{#1}}

\begin{document}
\title{Correlated electronic structure of the alternating monolayer-bilayer nickelate {L}a$_{5}${N}i$_{3}${O}$_{11}$}
\author{Harrison LaBollita}
\email{hlabollita@flatironinstitute.org}
\affiliation{Center for Computational Quantum Physics,
             Flatiron Institute,
             162 5th Avenue, New York, New York 10010, USA.}
\author{Antia S. Botana}
\affiliation{Department of Physics, Arizona State University, Tempe, AZ 85287, USA}

\begin{abstract}
The recent discovery of superconductivity under pressure in  Ruddlesden-Popper (RP) nickelates has attracted a great deal of attention. Here, using charge-self consistent density-functional theory plus dynamical mean-field theory, we study the correlated electronic structure of the latest superconducting member of the family: the alternating single-layer bilayer nickelate \lno{}. Due to its alternating single-layer and bilayer structural motif, this hybrid RP nickelate exhibits layer-selective physics with the single-layer neighboring a Mott instability, rendering the bilayer the dominant contributor to its low-energy physics, both at ambient and high pressure.  The electronic structure of \lno{} ultimately resembles that of the bilayer compound La$_{3}$Ni$_{2}$O$_{7}$, pointing to the presence of universal features in the family of superconducting RP nickelates. Thus, \lno{} provides a new platform to disentangle the key degrees of freedom underlying superconductivity in pressurized RP nickelates, underscoring the central role of the bilayer structural motif.
\end{abstract}

\maketitle

\section{\label{sec:intro}Introduction}
The recent reports of superconductivity in the pressurized bilayer La$_3$Ni$_2$O$_7$ ~\cite{sun2023superconductivity, wang2024bulk, hou2023emergence} and trilayer La$_4$Ni$_3$O$_{10}$ ~\cite{zhang2023superconductivity, li2023signature, zhu2024superconductivity} Ruddlesden-Popper (RP) nickelates reinvigorated the field after intensive exploration of the square-planar layered nickelates  ~\cite{Li2019superconductivity,Osada2020superconducting, Osada2021nickelate, Zeng2021superconductivity, Pan2021superconductivity}.  In the bilayer RP nickelate, superconductivity with a high critical temperature ($T_{c}$) $\sim$ 80 K has been experimentally determined in the bulk under pressures above 15 GPa ~\cite{sun2023superconductivity, wang2024bulk, hou2023emergence}. In the trilayer, a $T_c$ of $\sim20-30$ K can be achieved at a similar pressure threshold ~\cite{zhang2023superconductivity, li2023signature, zhu2024superconductivity} . In both cases, a structural transition concomitant with the onset of superconductivity takes place from an orthorhombic (bilayer) or monoclinic (trilayer) to a tetragonal crystal structure under pressure \cite{wang2023structure, zhu2024superconductivity}. This structural transition suppresses the tiltings of the NiO$_{6}$ octahedra present at ambient pressure. This series of discoveries established the RP nickelates R$_{n+1}$Ni$_n$O$_{3n+1}$ (R = rare
earth, $n$= number of perovskite-like layers along $c$) \cite{Greeenblatt1997ruddlesden} as a new family of superconductors. Formal valence counting gives an average Ni $3d$ filling of $d^{7+1/n}$, highlighting the increasingly important role of the $d_{3z^2-r^2}$ ($d_{z^2}$) orbitals -- an aspect that has drawn a plethora of theoretical attention and that introduced new puzzles in the general understanding of superconductivity in the nickelates ~\cite{zhang2023electronic, chen2023critical, lechermann2023electronic, christiansson2023correlated, luo2023bilayer, gu2023effective, shen2023effective, zhang2023electronic, wú2023charge, yang2023possible, liu2023swave, zhang2023structural, qu2023bilayer, yang2023minimal, zhang2023trends, lu2023superconductivity, tian2023correlation,vortex2023huang, jiang2023screening, liao2023correlations, liao2023interlayer,oh2023tj,qin2023singlets, sakakibara2023hubbard, shilenko2023, sakakibara2023theoretical,yi2024antiferromagnetic,labollita_stripes,huo2025,
zhu2025quantumphasetransitiondriven,
zhan2025,
le2025oppositemirrorparityscatteringoriginsuperconductivity,
lechermann2025lowenergyperspectiveinteractingelectrons,
maier2025interlayerpairingbilayernickelates,
lange2024,
jiang2025,
fan2024}.

To shed light on some of the open questions in the RP nickelates, expanding the search for superconductivity to other members of the family  has become a promising research direction. In this context, the new series of hybrid RP phases with general chemical formula R$_{n+1}$Ni$_n$O$_{3n+1}$ $\cdot$ R$_{m+1}$Ni$_m$O$_{3m+1}$ ($n$ $\neq$ m) is a prime target \cite{li2023design}. Here, rather than the uniform stacking of perovskite blocks found in RP phases, novel alternating stacking sequences with $m$ and $n$ blocks along $c$ are achieved instead. The single-layer trilayer (La$_{2}$NiO$_{4}\cdot$La$_{4}$Ni$_{3}$O$_{10}$) polymorph of La$_3$Ni$_2$O$_7$ (`1313') was the first hybrid RP nickelate to become superconducting under pressure with $T_c$ $\sim$ 80 K \cite{puphal2023unconventional, abadi2024electronic}.  More recently, superconductivity was also reported in the single-layer bilayer La$_{2}$NiO$_{4}\cdot$La$_{3}$Ni$_{2}$O$_{7}$ (\lno{}) hybrid RP (`1212') under pressure \cite{shi2025superconductivity}. 

In contrast to the conventional RP nickelates,
an orthorhombic phase without tilts
has been proposed for both of these hybrid phases at ambient pressure with pressure stabilizing a
tetragonal structure~\cite{chen2023polymorphism, puphal2023unconventional, li2023design, Wang2024longrange}. In La$_3$Ni$_2$O$_7$-1313, DFT+DMFT calculations concluded that the single-layer is
in a Mott insulating regime with the low-energy physics being dominated by the correlated trilayer block at ambient pressure and at high-pressures~\cite{lechermann2024electronic,labollita2024polym}. This electronic structure is consistent with recent angle-resolved photoemission spectroscopy (ARPES) measurements at ambient pressure~\cite{damascelli2025}. In contrast, constrained RPA calculations based on the non-magnetic DFT electronic structure showed that the leading pairing instability in both La$_3$Ni$_2$O$_7$-1313 and La$_5$Ni$_3$O$_{11}$-1212 is dominated by the single-layer block~\cite{zhang2024electronic, zhang2025electronic}. 

Motivated by the recent report of superconductivity in pressurized \lno{}~\cite{shi2025superconductivity}, we employ an advanced many-body electronic structure framework based on density-functional theory and dynamical mean-field theory to study the normal-state properties of this material at ambient and high-pressure. Our results reveal layer-selective physics regardless of pressure, with the bilayer block always dominating the low-energy physics. The evolution of the correlated electronic structure in this hybrid RP mimics that of the conventional bilayer RP nickelate. These findings not only identify \lno{}-1212 as a new platform for investigating superconductivity in hybrid RP nickelates but also emphasize the relevant role of the bilayer structural architecture in enabling-high $T_{c}$ superconductivity in the RP nickelate family.

\section{\label{sec:theory}Theoretical framework}
The charge self-consistent combination of density-functional theory (DFT) and dynamical mean-field theory (DMFT) is used to analyze the electronic structure of \lno{}-1212, where the Ni atoms serve as single-site quantum impurity problems. \edit{The DFT calculations are performed using an all-electron, full-potential framework built on an augmented plane-wave plus local orbital (APW+lo) basis set, as implemented in the \textsc{WIEN2}k code~\cite{Blaha2020wien2k} with the local-density approximation (LDA) for the exchange-correlation functional.} The basis set size is set by $R_{\mathrm{MT}}K_{\mathrm{max}}=7$ and muffin-tin radii (in a.u.) of 2.22, 1.80, and 1.60 for La, Ni, O, respectively. Brillouin zone integration is performed on a $7\times7\times2$ $\mathbf{k}$-grid.  For the DMFT part, local Ni($3d$) orbitals are obtained by projection from the Kohn-Sham (KS) bands in a large energy window spanning [$-10$, $5$] eV  around the Fermi level (\ef{}) which includes hybridization with the O($2p$), La($5d$), and La($4f$) states~\cite{Aichhorn2016dfttools,Aichhorn2009interface}. The three Ni sites in the primitive unit cell, each consist of a five $3d$-orbital correlated subspace in which a Slater Hamiltonian governs the interactions parameterized by a Hubbard $U=10$ eV and Hund's coupling $J_{\mathrm{H}} =1$ eV, \edit{which are reasonable values for the high energy hybridization considered~\cite{haule2014covalency} and are consistent with other nickelate calculations \cite{Pascut2023Correlation,lechermann2024electronic,christiansson2023correlated, Ouyang2025phase}}
\footnote{\edit{The two inequivalent Ni sites in La$_{5}$Ni$_{3}$O$_{11}$ stem from its bilayer and single layer structural components. In principle, the on-site Coulomb interacations could be different between them, however, the local Ni-O coordinations are similar enough that the screeened interactions are expected to lie within the same range. Ref. \onlinecite{Ouyang2025phase} employed two different sets of interaction parameters for the inequivalent Ni sites and obtained qualitatively similar results.}}. Each inequivalent single-site quantum impurity problem is solved using the continuous-time quantum Monte Carlo in the hybridization expansion as implemented in TRIQS/\textsc{cthyb}~\cite{Priyanka2016cthyb, Parcollet2015triqs}. The fully-localized limit (FLL) form of the double-counting correction is used~\cite{fll}. Real-frequency data are obtained from Matsubara data via analytic continuation. The local DMFT Green’s function is continued using the maximum entropy plus preblur method~\cite{Kraberger2017maxent} to obtain local $\mathbf{k}$-integrated spectra. The local self-enelgies are continued using the Pad\'e method to compute both $\mathbf{k}$-resolved and $\mathbf{k}$-integrated spectral functions from the Bloch Green’s function. A paramagnetic state is enforced in all calculations.

For the structural data, we employ the lattice constants at ambient pressure and at 15 GPa from Ref.~\onlinecite{shi2025superconductivity}. The ambient pressure crystal structure is orthorhombic with \textit{Cmmm} crystal symmetry, while the high-pressure (15 GPa) crystal structure is tetragonal with \textit{P4/mmm} crystal symmetry~\footnote{\edit{The internal atomic coordinates are optimized with a force convergence criterion of 1 mRy/au.}}.

\section{\label{sec:results}Results}

\begin{figure*}
    \centering
    \includegraphics[width=2\columnwidth]{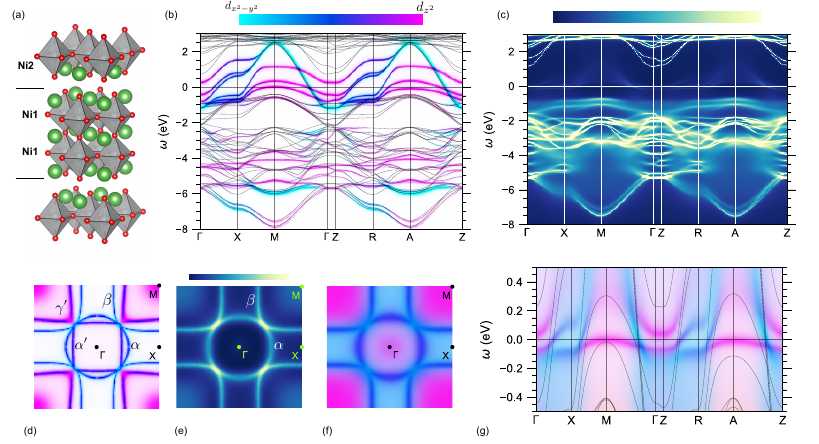}
    \caption{Structure and $\mathbf{k}$-dependent data of \lno{} at ambient pressure from DFT (b,d) and DFT+DMFT at $T\sim150$ K (c,e-g). (a) Crystal structure of \lno{} in the \textit{Cmmm} space group with inequivalent Ni sites denoted: Ni1 in the bilayer, and Ni2 in the single-layer block. (b) LDA band structure along high-symmetry lines in the Brillouin zone using the ``fatspectral'' representation of the Ni($3d$) orbitals. (c) False-color plot of the spectral function in a large energy window along high symmetry directions in the Brillouin zone. (d) LDA Fermi surface in the $k_{z}=0$ plane in the Ni($3d$) ``fatspectral'' representation. (e,f) Interacting Fermi surfaces in the $k_{z}=0$ plane: false-color plot (left) and Ni($3d$)-projected (right). (g) Low-energy blow up of the interacting spectral function in (c) using the Ni($3d$) fatspectral representation where thin lines denote the LDA band structure.}
    \label{fig:5311-0-Akw}
\end{figure*}

\textit{Crystal structure and DFT picture.--} 
\lno{} (referred to as the `1212' RP nickelate) is the $(n,m)=(1,2)$ member of the hybrid Ruddlesden-Popper series with general chemical formula La$_{n+1}$Ni$_{n}$O$_{3n+1}\cdot$La$_{m+1}$Ni$_{m}$O$_{3m+1}$ ($n\neq m$). Structurally, \lno{} consists of alternating perovskite-like layers of La$_{2}$NiO$_{4}$ ($n=1$) and La$_{3}$Ni$_{2}$O$_{7}$ ($m=2$) resulting in a 1-2-1-2 layering motif (see Fig.~\ref{fig:5311-0-Akw}(a)). At ambient pressure, the crystal structure is orthorhombic with \textit{Cmmm} crystal symmetry, containing two inequivalent Ni sites: Ni1 in the bilayer and Ni2 in the single-layer block.  As mentioned above, in contrast to the `conventional' bilayer nickelate La$_{3}$Ni$_{2}$O$_{7}$ (referred to as `2222'), the NiO$_{6}$ octahedra in \lno{} at ambient pressure exhibit no tilts as both the in-plane and out-of-plane Ni-O-Ni bond angles are 180$^{\circ}$. However, some structural ambiguities remain in the \lno{}-1212 compound as non-superconducting samples have also been resolved in the \textit{Immm} space group at ambient pressure~\cite{li2023design}. Formal valence counting gives an average Ni($3d$) occupation of $d^{7.67}$. For the structural constituents $n=1$ and $n=2$ RPs, the average Ni($3d$) occupations are $d^{8}$ and $d^{7.5}$, respectively. Hence, in terms of Ni($3d$) occupation, the hybrid RP nickelate represents an average of its $n=1$ and $n=2$ structural constituents.

Figure~\ref{fig:5311-0-Akw}(b) provides the non-magnetic DFT  description of the band structure of \lno{}-1212 using the ``fatspectral'' representation~\cite{lechermann2022} to emphasize the Ni($3d$) orbital content. Bands of Ni-$e_g$ character (hybridized with O(2$p$) states) from both the single-layer and bilayer blocks can be observed in the vicinity of the Fermi level (\ef{}): two Ni-$d_{x^{2}-y^{2}}$ bands from the bilayer and one from the single-layer block (blue), as well as one $d_{z^2}$ band from the single-layer and two Ni-$d_{z^{2}}$ from the bilayer (pink) -- the latter forming a bonding (B)-antibonding (AB) combination due to quantum confinement  \cite{Pardo2010quantum,jung2022electronic,labollita2023electronic}. The bonding Ni-$d_{z^{2}}$ sits just below \ef{}, while the anti-bonding Ni-$d_{z^{2}}$ band is $\sim 1$ eV above \ef{}. The Ni-$d_{z^{2}}$ band from the single-layer block sits in between the B-AB bands from the bilayer. The Ni-$t_{2g}$ bands span a 1 eV window between $-2$ and $-1$ eV relative to \ef{}. The corresponding Fermi surface at the DFT level in the $k_{z}=0$ plane is shown in Fig.~\ref{fig:5311-0-Akw}(d) where four Fermi surface sheets can be observed. Two sheets stem primarily from the bilayer Ni1 site: a circular electron-like pocket ($\alpha$) and a hole-pocket of mixed Ni-$e_{g}$/Ni-$d_{x^{2}-y^{2}}$ character ($\beta$). The remaining two sheets are contributed by the single-layer Ni2 site: a large hole pocket around the M point with Ni-$d_{z^{2}}$ character ($\gamma^{\prime}$) and  a square pocket at the zone center with Ni-$d_{x^{2}-y^{2}}$ character ($\alpha^{\prime}$). Thus, the electronic structure of \lno{}-1212  at the DFT level is quite different from that of the bilayer RP La$_3$Ni$_2$O$_7$-2222 where only the $\alpha$ and $\beta$ bands are present at the Fermi level \cite{zhang2023electronic,labollita2023electronic}.  

As shown in previous work \cite{lechermann2024electronic,labollita2024polym}, a non-magnetic DFT description may not be sufficient to provide a faithful description of the electronic structure of the existing hybrid RP nickelates, as it cannot account for a Mott-insulating state in the single-layer La$_2$NiO$_4$. This conclusion is particularly obvious in the single-layer trilayer polymorph of La$_{3}$Ni$_{2}$O$_{7}$-1313. In this material, the electronic structure at the non-magnetic DFT level also shows extra Fermi surface sheets coming from the single-layer block \cite{chen2023polymorphism,labollita2024polym}. However, recent ARPES experiments do not show any evidence for these extra sheets \cite{damascelli2025}. The electronic structure observed in ARPES can instead be easily captured by including beyond DFT electronic correlations: in DFT+DMFT the single-layer block is indeed gapped (in line with the Mott insulating state of bulk La$_2$NiO$_4$ \cite{cava1991}), and the only contributions at the Fermi level become those from the trilayer block, in agreement with ARPES. With these considerations, we now turn to the role of local electronic correlations in \lno{}-1212 using DFT+DMFT.

\begin{figure}
    \centering
    \includegraphics[width=\columnwidth]{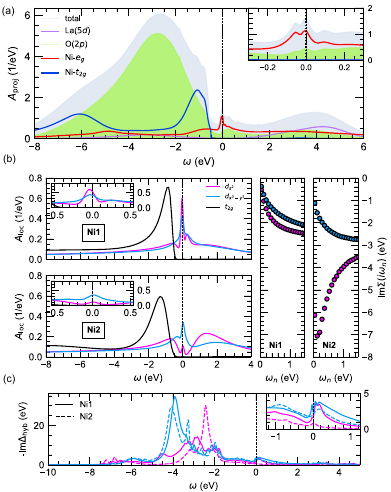}
    \caption{Total- and site-resolved $\mathbf{k}$-integrated spectral data of \lno{} at ambient pressure (\textit{Cmmm}) at $T\sim 150$ K. (a) Orbital-projected spectral function (inset: low-energy blow up). (b) Left panel: site-resolved local Ni($3d$) spectral functions (inset: low-energy blow up). Right panel: site-resolved local Ni-$e_{g}$ imaginary part of the Matsubara self-energies. (c) Site-resolved Ni-$e_{g}$ hybridization functions (inset: low-energy blow up).}
    \label{fig:5311-0-Aproj}
\end{figure}

\textit{Correlated electronic structure at ambient pressure.--}
Figure~\ref{fig:5311-0-Akw}(c,g) shows the $\mathbf{k}$-resolved spectral data for  \lno{} at ambient pressure from DFT+DMFT at $T\sim 150$ K ($\beta = 1/T = 77$ eV$^{-1}$). The Ni-$e_{g}$ dominated parts of the spectrum are very strongly correlated with large scattering rates rates for both Ni1 (bilayer) and Ni2 (single-layer), which results in heavily renormalized and decoherent Ni-$e_{g}$ dispersions in both orbital sectors (see below). For the Ni2 site, the $d_{z^{2}}$ orbital develops a Mott gap, while the $d_{x^{2}-y^{2}}$ orbital retains finite low-energy weight but with a large scattering rate (see Fig.~\ref{fig:5311-0-Aproj}(b)). Taken together, this yields ``bad metal'' behavior, proximate to a Mott insulating state for the Ni2 site, while the dominant low-energy spectral weight arises from Ni1-$e_{g}$ states, in contrast to the band theory picture. The site- and orbital-projected DFT+DMFT momentum-resolved spectral functions confirm this description (see Fig. \ref{fig:site-proj-akw} of Appendix \ref {app:dmft}1). Coherent low-energy states are clearly dominated by Ni1-$e_{g}$ character, consistent with our description of the Ni1 site as metallic. By contrast, the Ni2 site contributes only incoherent spectral weight, primarily from the $d_{x^{2}-y^{2}}$ orbital, in line with its bad-metallic character proximate to a Mott insulator. The Ni2-$d_{z^{2}}$ projection highlights the suppression of this orbital's weight at the chemical potential consistent with the opening of a gap. This scenario aligns with previous DFT+DMFT calculations using smaller Slater integrals ($F_{k}$) to parameterize the Coulomb tensor~\cite{Ouyang2025phase}.

Similarly to the results in La$_3$Ni$_2$O$_7$-1313 \cite{lechermann2024electronic,labollita2024polym}, DFT+DMFT calculations in \lno{}-1212 reveal a much simpler electronic structure relative to that of DFT as only the bilayer contributes to the low-energy physics with two Ni1-$d_{x^{2}-y^{2}}$ bands (blue color) and the flat bonding Ni1-$d_{z^{2}}$ (pink color) remaining around \ef{}. The spectral weight corresponding to the anti-bonding Ni1-$d_{z^{2}}$ is renormalized to $+0.3$ eV above \ef{}, while the Ni-$t_{2g}$ dispersions remain below the chemical potential. The interacting Fermi surface in Fig.~\ref{fig:5311-0-Akw}(e,f) clearly reflects the simplified electronic structure obtained at the DFT+DMFT level for \lno{}-1212. Only two Fermi surface sheets can now be observed: the circular electron pocket ($\alpha$) and the large hole-like pocket ($\beta$), both with predominant Ni-$d_{x^{2}-y^{2}}$ orbital character. The effect of local dynamical correlations is to remove the Fermi surface sheets contributed by the single-layer block, leaving only the bilayer sheets. Thus, by incorporating electronic correlations, we find an emergent electronic structure in \lno{}-1212 that is nearly identical to that of La$_{3}$Ni$_{2}$O$_{7}$-2222 at ambient pressure~\cite{Yang2024Orbital,damascelli2025}, as obtained from ARPES experiments.

\begin{figure}
    \centering
    \includegraphics[width=\columnwidth]{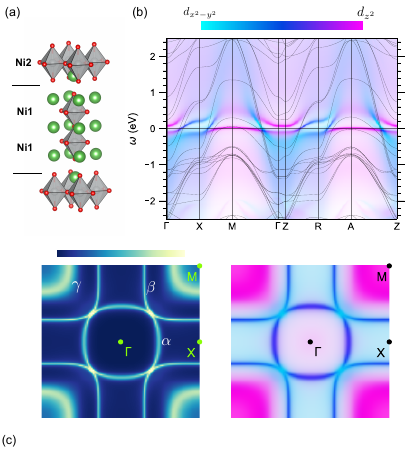}
    \caption{$\mathbf{k}$-dependent spectral data of \lno{}  at 15 GPa  (\textit{P4/mmm}) from DFT+DMFT at $T\sim 116$ K. (a) Crystal structure of \lno{} in the \textit{P4/mmm} space group with inequivalent Ni sites labeled. (b) Interacting spectral function in Ni($3d$) fatspectral representation along high-symmetry lines. Thin, black lines correspond to the LDA band structure. (c) Interacting Fermi surface in the $k_{z}=0$ plane: false-color plot (left) and Ni($3d$)-projected (right).}
    \label{fig:5311-15GPa-Akw}
\end{figure}

To further elucidate correlation effects on the electronic structure of \lno{}-1212, we analyze the site-resolved local Ni($3d$) $\mathbf{k}$-integrated spectral functions and Matsubara Ni-$e_{g}$ self-energies shown in Fig.~\ref{fig:5311-0-Aproj}(b). The local spectral function on the Ni1 site corresponding to the bilayer reveals a broad Ni-$d_{z^{2}}$ and Ni-$d_{x^{2}-y^{2}}$ spectral weight, indicating decoherence in both orbital sectors, as mentioned above.c

The site-resolved Ni-$e_{g}$ Matsubara self-energies are shown in the right panel of Fig.~\ref{fig:5311-0-Aproj}(b). From the low-Matsubara behavior of the self-energy, we can determine key features of the correlated electronic structure. As a metric for the strength of electronic correlations, the mass enhancement $m^{\star}/m_{\mathrm{DFT}} = 1-\partial_{\omega_{n}}\mathrm{Im}\Sigma(i\omega_{n})|_{i\omega_{n}\rightarrow0}$ is often used. Fitting a fourth order polynomial ($\mathrm{Im}\Sigma(i\omega_{n}>0) \approx \sum_{p}a_{p}\omega_{n}^{p}$) to the lowest six Matsubara frequencies yields $m^{\star}_{d_{z^{2}}}\sim10$ and $m^{\star}_{d_{x^{2}-y^{2}}}\sim8$ for the bilayer Ni (Ni1), which are comparable to bulk ambient pressure La$_{3}$Ni$_{2}$O$_{7}$-2222~\cite{lechermann2023electronic}. The scattering rate ($\propto -\mathrm{Im}\Sigma(i\omega_{n})|_{i\omega_{n}\rightarrow 0}$) on both Ni-$e_{g}$ orbitals remains large at this temperature underpinning the `washed out' spectral features in the $\mathbf{k}$-dependent data rendering the bilayer in a bad metal or incoherent metallic state. The Ni2-$e_{g}$ self-energies are close to a Mott instability with a diverging Ni-$d_{z^{2}}$ component and large Ni-$d_{x^{2}-y^{2}}$ component.

\begin{figure}
    \centering
    \includegraphics[width=\columnwidth]{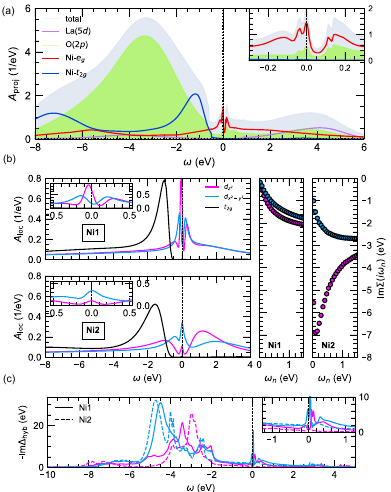}
    \caption{Total- and site-resolved $\mathbf{k}$-integrated spectral data of \lno{} at 15 GPa (\textit{P4/mmm}) at $T\sim 116$ K. (a) Orbital-projected spectral function (inset: low-energy blow up). (b) Left panel: site-resolved local Ni($3d$) spectral functions (inset: low-energy blow up). Right panel: site-resolved local Ni-$e_{g}$ imaginary part of the Matsubara self-energies. (c) Site-resolved Ni-$e_{g}$ hybridization functions (inset: low-energy blow up).}
    \label{fig:5311-15GPa-Aloc}
\end{figure}

Connecting the local spectrum back to the Bloch spectrum, we analyze the DMFT hybridization function $\Delta_{\mathrm{hyb}}(\omega) = \omega+\mu - G^{-1}_{\mathrm{loc}}(\omega) - \Sigma_{\mathrm{imp}}(\omega)$. The hybridization function $-\mathrm{Im}\Delta_{\mathrm{hyb}}$ shown in Fig.~\ref{fig:5311-0-Aproj}(c) reaffirms the distinct orbital- and site-selective behavior. For the Ni1 site, both orbitals exhibit large finite values of $\mathrm{Im}\Delta(\omega=0)$, indicative of strong hybridization with the bath and itinerant electrons. In contrast, the Ni2 site shows orbital differentiation (in line with the Ni2 self-energies): the Ni-$d_{z^{2}}$  orbital exhibits vanishing hybridization at low frequencies, signaling strong localization and effective decoupling from the bath, while the Ni-$d_{x^{2}-y^{2}}$ orbital retains a finite, though significantly reduced, hybridization, pointing to weak coupling to metallic states.  Combining this analysis with the Matsubara self-energies, the data support a picture of orbital- and site-selective localization, where Ni2 hosts a Mott-localized orbital and a weakly hybridized, incoherent metallic orbital, while Ni1 exhibits itinerancy without coherence.

To scrutinize orbital admixture effects on the electronic structure of \lno{}-1212 at ambient pressure, we carefully examine the total and orbital-projected $\mathbf{k}$-integrated spectral data including the O(2$p$) states (see Fig.~\ref{fig:5311-0-Aproj}(a)). The sizeable admixture of Ni-$e_{g}$ and O($2p$) content can be observed. Furthermore, the spectral features include broad O($2p$) peak at $-3.5$ eV and a Ni($3d$) peak $-0.5$ eV. The dominant $d$ and $p$ peaks can be translated into a charge-transfer energy $\Delta_{\mathrm{CT}}=\varepsilon_{d}-\varepsilon_{p}\sim3$ eV. We observe only minor La($5d$) weight located within the broader O($2p$) peak. Total Ni($3d$) occupations read $\sim 8.2$ for both Ni sites indicating sizeable hole character on the oxygens~\cite{lechermann2022assessing,lechermann2023electronic,lechermann2024electronic} (see Table \ref{tab:occ} of Appendix \ref{app:dmft}3). Ultimately, in terms of filling, both Ni sites exhibit nearly half-filled Ni-$e_{g}$ orbitals~\cite{lechermann2023electronic,leonov2023correlated}. To summarize, the DFT+DMFT electronic structure of \lno{}-1212 at ambient pressure  is very similar to that of La$_3$Ni$_2$O$_7$-2222 as shown in the spectral data. We subsequently analyze the evolution of the electronic structure of the single-layer bilayer RP nickelate  under pressure.

\textit{Correlated electronic structure at 15 GPa.--} 
\lno{}-1212 becomes superconducting when moderate pressures are applied ($\sim10-25$ GPa) \cite{shi2025superconductivity}. The emergence of superconductivity is concomitant with the suppression of a density-wave-like state and a structural transition from an orthorhombic to a tetragonal space group symmetry (see Fig.~\ref{fig:5311-15GPa-Akw}(a))~\cite{shi2025superconductivity},  akin to its bilayer counterpart~\cite{liu2023evidencesdw,chen2023musr,Khasanov2024musr,kakoi2024nmr, chen2024rixs,xie2024neutrons}.  The momentum-resolved spectral function for \lno{} at 15 GPa from DFT+DMFT  at $T\sim 116$ K is shown in Fig.~\ref{fig:5311-15GPa-Akw}(b) within the fatspectral representation.  Increased electronic bandwidths can be observed as a consequence of the the reduction in lattice constants (of $\sim2$\%) and the related contraction of NiO$_{6}$ octahedra. The combination of pressure and lower temperature increase the coherence giving rise to quasiparticle (QP) dispersions. The key low-energy dispersions are once again of Ni-$e_{g}$ orbital content stemming exclusively from the bilayer. The most relevant effect of pressure is the pinning of the bonding Ni-$d_{z^{2}}$ dispersion at \ef{}. As a consequence, the interacting Fermi surface (shown in Fig.~\ref{fig:5311-15GPa-Akw}(c)) now shows a third pocket of $d_{z^2}$ character at the zone corners ($\gamma$) -- in addition to the $\alpha$ and $\beta$ sheets obtained at ambient pressure. While studying the superconducting pairing symmetry of \lno{}-1212 would require a tailored model Hamiltonian study (which is beyond the scope of this work), we note that the appearance of the $\gamma$ corner pocket in bulk La$_{3}$Ni$_{2}$O$_{7}$ under pressure has been linked to the emergence of superconductivity ~\cite{zhang2023structural,liu2023swave,gu2023effective,qu2023bilayer,yang2023possible,yang2023minimal,liao2023interlayer}.

The $\mathbf{k}$-integrated spectral data and site-resolved local quantities are summarized in Fig.~\ref{fig:5311-15GPa-Aloc}. The total- and orbital-resolved spectral function in Fig.~\ref{fig:5311-15GPa-Aloc}(a) reveals broad O($2p$) and Ni($3d$) peaks at  $\sim$ $-4$ eV  and $-0.75$ eV, respectively, resulting in a slightly increased charge-transfer energy relative to ambient pressure~\cite{labollita2023electronic}. The metallic spectral weight is dominantly Ni-$e_{g}$ in character with some admixture of O($2p$) content. A sharp peak at \ef{} has emerged due to the increased coherence. There is no contribution from the Ni-$t_{2g}$ orbitals around \ef{}. The effect of pressure on the local Ni sites is subtle (see Fig.~\ref{fig:5311-15GPa-Aloc}(b)), and the coherent low-energy slates are still dominated by Ni1-$e_g$ character (see also Fig. \ref{fig:site-proj-akw} of Appendix \ref{app:dmft}2). As pressure increases the electronic bandwidth, electronic correlations are weakened for both Ni sites, as evidenced by the self-energies in Fig.~\ref{fig:5311-15GPa-Aloc}(b), where the reduced mass enhancements are $m_{z^{2}}$ $\sim$ 6  and $m_{x^{2}-y^{2}}$ $\sim$ 4. In addition to the reduced mass enhancements, the scattering rates have also decreased, consistent with the increased coherence of both orbitals in the bilayer. Connecting the total spectrum to the local spectrum, we find that the sharp QP-like peak in the total spectrum can be assigned to the Ni-$d_{z^{2}}$ orbital within the bilayer associated with the flat-band in the $\mathbf{k}$-resolved spectral function (see Fig.~\ref{fig:5311-15GPa-Akw}(b)). This coincides with the reduced Im$\Delta_{\mathrm{hyb}}$ for Ni1-$d_{z^{2}}$ (See Fig.~\ref{fig:5311-15GPa-Aloc}(c)). Additionally, the effect of pressure has transformed the broad Ni-$d_{x^{2}-y^{2}}$ spectral weight at ambient pressure into a subtle QP-like peak with large hybridization function at \ef{} denoting itinerant character. We emphasize the two key features of the pressurized electronic structure of \lno{}: the absence of contributions from the single-layer block to the low-energy electronic structure, and the emergence of the $\gamma$ pocket at the zone corners, which is consistent with pressurized La$_{3}$Ni$_{2}$O$_{7}$-2222.

\section{\label{sec:summary} Summary and discussion}
By combining density-functional theory and dynamical mean-field theory, we have identified several relevant features in the electronic structure of the single-layer bilayer hybrid RP nickelate \lno{} at both ambient and high pressure. Incorporating electronic correlations reveals layer-selective behavior in \lno{}, as the single-layer is effectively localized -- either in or near a Mott insulating state -- while the bilayer remains metallic but strongly correlated. Thus, our results suggest that the low-energy physics of this hybrid RP nickelate is dominated by the bilayer block while the single-layer acts primarily as a `spectator' and is unlikely to drive the observed superconductivity. As a consequence, the electronic structure of \lno{}-1212 is very similar to that of the bilayer RP La$_3$Ni$_2$O$_7$-2222, suggesting that they are susceptible to the same low-energy phenomena. In both materials two Fermi surface sheets are present at ambient pressure ($\alpha$ and $\beta$) of dominant $d_{x^2-y^2}$ character. Under pressure, an extra corner pocket $\gamma$ of pure $d_{z^2}$ character arises that has been linked to the emergence of superconductivity in the bilayer RP. Within the metallic bilayer, orbital-selective differentiation emerges: the flat Ni-$d_{z^{2}}$ orbitals show strong correlations and a high density of states, while the Ni-$d_{x^{2}-y^{2}}$ are also correlated but more itinerant. The striking similarities in the electronic structure of La$_{3}$Ni$_{2}$O$_{7}$-2222 and \lno{}-1212 (despite obvious differences in the structure and electron count) suggest an important role of the bilayer structural motif and universal features in the electronic structure necessary for superconductivity to arise in RP nickelates.

Future interesting directions include ARPES and optical spectroscopy as probes of the electronic structure of \lno{}-1212 at ambient pressure and tests of the theory presented here. Given the exciting developments in the thin-film community on strained La$_{3}$Ni$_{2}$O$_{7}$~\cite{Ko2025Signatures} and variants~\cite{liu2025superconductivity} we envision that strain-induced superconductivity in this hybrid RP will also be an interesting avenue to explore both experimentally and theoretically. Furthermore, it will be promising to span the studies of correlated phenomena in higher-order $(n,m)$ hybrid RP nickelates.

\section*{Acknowledgments}
ASB was supported by NSF grant No. DMR-2045826. The Flatiron Institute is a division of the Simons Foundation.

\appendix

\begin{figure*}
\centering
\includegraphics[width=1.75\columnwidth]{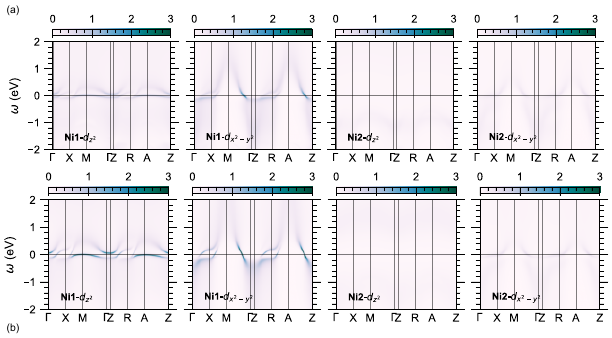}
\caption{Site- and orbital-projected DFT+DMFT momentum-resolved spectral functions along high-symmetry lines for \lno{} at (a) ambient pressure and (b) 15 GPa.}
\label{fig:site-proj-akw}
\end{figure*}

\begin{figure}[h!]
    \centering
    \includegraphics[width=\columnwidth]{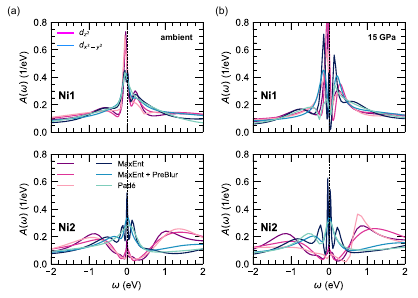}
    \caption{Site- and orbital-resolved local DFT+DMFT spectral functions of for La$_{5}$Ni$_{3}$O$_{11}$ using the MaxEnt (solid lines) and Pad\'e (dashed lines) analytic continuation methods at (a) ambient pressure and (b) 15 GPa.}
    \label{fig:pade}
\end{figure}

\section{\label{app:dmft}Additional DFT+DMFT data} 
In this appendix, we provide additional DFT+DMFT data for \lno{} at ambient pressure and 15 GPa. Sec.~\ref{app:akw} discusses additional spectral function data. Sec~\ref{app:ac} compares maximum entropy and Pad\'e analytic continuation for the local spectral functions. Sec~\ref{app:occ} provides the DFT+DMFT occupations.

\subsection{\label{app:akw}Site- and orbital-resolved spectral function}
To further clarify the site- and orbital character of the low-energy states in the interacting spectrum of \lno{}, we project the spectral function $A(\mathbf{k},\omega)$ onto individual Ni sites and Ni($3d$) orbitals, yielding $A^{\mathrm{Ni}_{1}}_{mm'}(\mathbf{k},\omega)$ and $A^{\mathrm{Ni}_{2}}_{mm'}(\mathbf{k},\omega)$, where $m,m'$ are Ni($3d$) orbital indices (see Figs.~\ref{fig:site-proj-akw}). These data provide complementary information to the main text figures and allow us to disentangle the contributions of inequivalent sites and orbitals to the low-energy electronic structure.

In both ambient and high-pressure calculations, the coherent low-energy states are dominated by Ni1-$e_{g}$ character, consistent with our description of the Ni1 site as metallic. By contrast, the Ni2 site contributes only incoherent spectral weight, primarily from the $d_{x^{2}-y^{2}}$ orbital, in line with its bad-metallic character proximate to a Mott insulator. The Ni2-$d_{z^{2}}$ projection highlights the suppression of this orbital weight at the chemical potential, consistent with the opening of a gap.

Together, these site- and orbital-resolved data emphasize the site-selective correlations in \lno{} and illustrate the dominant low-energy spectral weight arises from the Ni1-$e_{g}$ states rather than Ni2, in stark contrast to band theory expectations.

\subsection{\label{app:ac}Comparison between MaxEnt and Pad\'e analytic continuation}
In the main text, we used the maximum entropy (MaxEnt) plus preblur method to obtain real-frequency spectra from the local imaginary-time Green's functions. MaxEnt can generate artificial sharp structures in metallic systems because of its regularization procedure. To verify the robustness of our conclusions, we also perform analytic continuation using Pad\'e approximates and compare the two methods.

We start by comparing the spectra obtained with the MaxEnt (without preblur) to the Pad\'e methods, respectively in Fig.~\ref{fig:pade}. The comparison shows that the gross features of the spectra are consistent, but there are important qualitative differences. In particular, the sharp peaks that appear in the Ni2-$d_{x^{2}-y^{2}}$ spectrum when using MaxEnt are absent with Pad\'e, which instead yields broad, incoherent spectral weight. Likewise, the three peak structure suggested by MaxEnt does not appear with Pad\'e. These differences indicate that the sharp low-energy features in the MaxEnt spectra are numerical artifacts rather than physical. However, with the inclusion of the preblur technique (see Ref.~\cite{Kraberger2017maxent} for more details), we can obtain qualitative consistent spectra at both low- and high-energies. The preblur method takes as an input the standard deviation of the Gaussian distribution which we take to be 0.05.

Taken together, the two continuation methods present a clear picture: the Ni1 site provides the dominant coherent spectral weight at the Fermi level, while the Ni2 site remains incoherent with broadened spectral weight. This comparison supports the site-selective interpretation presented in the main text.

\begin{table}[h!]
\centering
\begin{tabular*}{\columnwidth}{c|@{\extracolsep{\fill}}cccc}
\hline\hline
  Ambient  & $d_{z^{2}}$ & $d_{x^{2}-y^{2}}$ & $t_{2g}$ & $3d$ total\\
  Ni1      &    1.15     &    1.10           & 5.96     &  8.21\\
  Ni2      &    1.14    &    1.10           & 5.97     &   8.21\\
\hline
  15 GPa  & $d_{z^{2}}$ & $d_{x^{2}-y^{2}}$ & $t_{2g}$ & $3d$ total\\
  Ni1      &    1.17     &     1.10          &    5.95  & 8.22\\
  Ni2      &    1.16     &     1.13          &    5.94  & 8.23 \\
\hline\hline
\end{tabular*}
\caption{Orbital-resolved DFT+DMFT occupations for La$_{5}$Ni$_{3}$O$_{11}$ at ambient pressure and 15 GPa.}
\label{tab:occ}
\end{table}

\subsection{\label{app:occ} Orbital Occupations}
The occupation data from our DFT+DMFT calculations for \lno{}-1212 at ambient pressure and at 15 GPa is summarized in Table~\ref{tab:occ}. We find that the overall Ni($3d$) occupations are similar.

\bibliography{ref.bib}

\end{document}